\documentclass[onecolumn,11pt]{article}%
\usepackage[utf8]{inputenc}
\usepackage{graphicx}
\usepackage{amsmath}
\usepackage{cite}%
\usepackage{amsfonts}%
\usepackage{amssymb}
\newcommand \be{\begin{equation}}
\newcommand \ee{\end{equation}}

\begin{document}

\title{Exact Microcanonical Formulation and Thermodynamics of Equispaced Finite-Level Systems}
\author{J. Ricardo de Sousa\\Universidade Federal do Amazonas,\\Departamento de F\'{\i}sica, \\3000, Japiim, 69077-000, Manaus-AM, Brazil}
\maketitle

\begin{abstract}
We present an exact microcanonical formulation, in the thermodynamic limit,
for a system of $N$ noninteracting particles with $p$ equally spaced energy
levels $\{0,\varepsilon,2\varepsilon,\ldots,(p-1)\varepsilon\}$. Writing the
microcanonical multiplicity $\Omega_{p}(E,N)$ as the coefficient of a
generating function and evaluating the resulting representation by
saddle-point analysis, we derive analytical expressions for the entropy per
particle $s(u,p)$ and inverse temperature $\beta(u,p)$, with
$u=E/(N\varepsilon)\in\lbrack0,p-1]$. The formulation applies to arbitrary $p$
and recovers the known cases $p=2$, $p=3$, and $p\rightarrow\infty$. For
finite $p$, the bounded spectrum implies an entropy maximum at $u_{c}%
=(p-1)/2$, where $\beta$ vanishes and changes sign. In the limit
$p\rightarrow\infty$, the upper spectral bound is lost, the finite-energy
entropy maximum disappears, and no negative-temperature branch remains. To our
knowledge, this is the first general thermodynamic-limit microcanonical
solution for arbitrary $p$. It therefore provides a unified framework for the
thermodynamics of equispaced finite-level systems and their bounded-spectrum
crossover with increasing $p$.

\end{abstract}

The microcanonical thermodynamics of systems with a finite number of equally
spaced energy levels provides a natural setting for the study of bounded
spectra. For finite $p$, the spectrum is bounded from above, so that the
entropy may attain a maximum at finite energy; at this point the inverse
temperature vanishes, and its subsequent sign change marks the continuation to
the negative-temperature branch\cite{r1,r2}. This thermodynamic structure is
well known in principle and has long been associated with bounded-spectrum
systems, including the classic nuclear-spin realization of Purcell and
Pound\cite{r3}. More broadly, the characterization of entropy and temperature
in isolated systems remains a topic of continuing interest in statistical
mechanics\cite{r4,r5}. In this work, we consider the energy spectrum%

\begin{equation}
\{0,\varepsilon,2\varepsilon,....,(p-1)\varepsilon\},\label{1}%
\end{equation}
with $\varepsilon>0$. This spectrum is equivalently described by a system of
$N$ noninteracting particles, each of which can occupy one of the $p$ local
states specified by Eq. (\ref{1}). The corresponding Hamiltonian may be
written as%

\begin{equation}
\mathcal{H}=\varepsilon\sum\limits_{i=1}^{N}\sigma_{i},\text{
\ \ \ \ \ \ \ \ \ \ \ }\sigma_{i}=0,1,2,...,(p-1).\label{1a}%
\end{equation}

Some particular cases of this problem are already known. These include the
standard two-level system ($p=2$), the three-level case ($p=3$) discussed
explicitly in Ref. \cite{liao}, and the limit $p\rightarrow\infty$. In these
cases, explicit expressions for the entropy per particle and the inverse
temperature can be obtained. To the best of our knowledge, however, a general
microcanonical treatment for arbitrary $p$ has not been presented in the
thermodynamic limit. In this respect, the three-level result is especially
relevant, since it provides a nontrivial finite-$p$ benchmark against which
the general formulation can be tested.

A general solution for arbitrary $p$ is desirable for at least two reasons.
First, it places the known cases $p=2$, $p=3$, and the limit $p\rightarrow
\infty$ within a single unified framework. Second, it makes it possible to
determine explicitly how the thermodynamic functions evolve as the number of
accessible energy levels increases. In particular, one can follow how the
entropy changes with energy, how the location of its maximum depends on $p$,
and how the corresponding temperature behavior evolves from finite bounded
spectra to the infinite-level limit. Moreover, since equally spaced finite
spectra arise naturally, for example, in spin systems with a finite Zeeman
multiplet, the general treatment is relevant not only as a formal extension of
known special cases, but also as a useful framework for the microcanonical
analysis of physically realizable bounded-spectrum systems.

In this Letter, we present an exact thermodynamic-limit microcanonical
treatment of the model for arbitrary $p$. Using a generating-function
formulation combined with saddle-point analysis, we derive analytical
expressions for the entropy per particle and the inverse temperature,
recovering as particular limits the known results for $p=2$, $3$, and
$p\rightarrow\infty$.

To introduce the method in its simplest nontrivial finite setting, and to
provide a direct benchmark against the available three-level solution, we
first consider the case $p=3$, with single-particle spectrum
\begin{equation}
\{0{\footnotesize ,\ }\varepsilon{\footnotesize ,}\text{ }2\varepsilon
\}.\label{2}%
\end{equation}
Let $N_{1}$, $N_{2}$, and $N_{3}$ denote the occupation numbers of the three
levels. For a macrostate with $N$ particles and total energy $E$, these
variables satisfy%

\begin{equation}
N=N_{1}+N_{2}+N_{3},\text{ \ \ \ \ \ }E=\varepsilon\left(  N_{2}%
+2N_{3}\right)  .\label{3}%
\end{equation}
Writing $q=E/\varepsilon$, one may take $N_{3}$ as the independent variable,
so that%
\begin{equation}
N_{2}=q-2N_{3},\text{ \ \ \ \ \ \ \ \ \ \ \ \ \ }N_{1}=N-q+N_{3}.\label{4}%
\end{equation}

For each allowed value of $N_{3}$, the corresponding number of configurations
is given by the multinomial factor%
\begin{equation}
\frac{N!}{N_{1}!N_{2}!N_{3}!}.\label{5}%
\end{equation}

Summing over all admissible values of $N_{3}$, one obtains%

\begin{equation}
\Omega_{3}(E,N)=\sum\limits_{r=r_{0}}^{\left[  q/2\right]  }\frac{N!}%
{r!\left(  N-q+r\right)  !\left(  q-2r\right)  !},\label{6}%
\end{equation}
where $r=N_{3}$ and $r_{0}=\max\left(  0,q-N\right)  $.

Although the thermodynamics of the three-level case has been discussed
previously in the literature, the present formulation is different in spirit,
since it is introduced here as the natural starting point for a
generating-function treatment that extends directly to arbitrary $p$. In
particular, the same counting can be written more compactly in terms of the
generating function%

\begin{equation}
\Omega_{3}(E,N)=\left[  x^{q}\right]  \left(  1+x+x^{2}\right)  ^{N},\label{9}%
\end{equation}
with $q=0,1,2,...,2N$. Here, $\left[  x^{q}\right]  f(x)$ denotes the
coefficient of $x^{q}$ in the expansion of $f(x)$, so that the coefficient of
$x^{q}$ in $\left(  1+x+x^{2}\right)  ^{N}$ gives the number of configurations
with total energy $E=q\varepsilon$.

Written in this form, the extension to an arbitrary number $p$ of equally
spaced levels becomes immediate. For the spectrum defined in Eq. (\ref{1}),
the number of microstates with total energy $E=q\varepsilon$ is%

\begin{equation}
\Omega_{p}(E,N)=\left[  x^{q}\right]  \left(  1+x+x^{2}+....+x^{p-1}\right)
^{N},\label{11}%
\end{equation}
with $q=0,1,2,...,(p-1)N$. Since the polynomial in Eq. (\ref{11}) is a finite
geometric sum, the microstate counting can be rewritten in the compact form%
\begin{equation}
\Omega_{p}(E,N)=\left[  x^{q}\right]  \left(  \frac{1-x^{p}}{1-x}\right)
^{N}.\label{13}%
\end{equation}
This reformulation is not merely algebraic. It provides the key step that
makes the arbitrary$-p$ problem analytically tractable in the thermodynamic
limit $N\rightarrow\infty$, since it converts the combinatorial counting
problem into a form that can be treated systematically by complex-analytic
methods. The factor $\left(  1-x^{p}\right)  /\left(  1-x\right)  $ also
coincides formally with the one-particle canonical partition sum under the
identification $x=e^{-\beta\varepsilon}$. In the present context, however, it
appears as a generating function, and the coefficient extraction $\left[
x^{q}\right]  $ enforces the fixed-energy constraint. The derivation therefore
remains genuinely microcanonical throughout. More broadly, the present
construction exemplifies a general generating-function principle for discrete
systems with additive observables, in which canonical partition functions
generate refined microcanonical multiplicities and the latter are recovered by
coefficient extraction.

Indeed, the coefficient extraction $\left[  x^{q}\right]  $ can be written as
a contour integral by means of Cauchy's formula. If%

\begin{equation}
f(z)=\sum\limits_{n=0}^{\infty}a_{n}z^{n},\label{14}%
\end{equation}
then%
\begin{equation}
a_{n}=\frac{1}{2\pi i}\oint\frac{f(z)}{z^{n+1}}dz.\label{15}%
\end{equation}

Applying this result to Eq. (\ref{13}), we obtain%
\begin{equation}
\Omega_{p}(E,N)=\frac{1}{2\pi i}\oint\frac{dz}{z^{q+1}}\left(  \frac{1-z^{p}%
}{1-z}\right)  ^{N}.\label{16}%
\end{equation}
Equation (\ref{16}) may therefore be regarded as the corresponding discrete
inversion formula, recovering the microcanonical multiplicities at fixed
energy from the generating function.

To extract the thermodynamic behavior, it is convenient to introduce the
reduced energy per particle,%
\begin{equation}
u=\frac{E}{N\varepsilon}=\frac{q}{N},\label{17}%
\end{equation}
so that $q=Nu$. Equation (\ref{16}) can then be cast in the exponential form%
\begin{equation}
\Omega_{p}(E,N)=\frac{1}{2\pi i}\oint\frac{dz}{z}e^{N\phi_{p}(z)},\label{18}%
\end{equation}
where
\begin{equation}
\phi_{p}(z)=\ln\left(  \frac{1-z^{p}}{1-z}\right)  -u\ln z.\label{19}%
\end{equation}
Equation (\ref{18}) is the central representation of the problem, since it
gives direct access to the thermodynamic limit $N\rightarrow\infty$. Its
contour-integral form and asymptotic evaluation follow standard methods of
coefficient extraction and saddle-point analysis\cite{mec,r6}. In this limit,
the dominant contribution is determined by the saddle-point condition%

\begin{equation}
\phi_{p}^{\prime}(z_{0})=0,\label{20}%
\end{equation}
which yields%
\begin{equation}
u=\frac{z_{0}}{1-z_{0}}-\frac{pz_{0}^{p}}{1-z_{0}^{p}}.\label{21}%
\end{equation}

If $x=x(u)$ denotes the physically relevant solution of Eq. (\ref{21}), the
Boltzmann entropy per particle in the thermodynamic limit is given by%

\begin{equation}
s(u,p)=\lim_{N\rightarrow\infty}\left[  \frac{S_{p}(E,N)}{N}\right]
=k_{B}\left[  \ln\left(  \frac{1-x^{p}}{1-x}\right)  -u\ln(x)\right]
.\label{22}%
\end{equation}

Equation (\ref{22}) provides the fundamental thermodynamic relation of the
model for arbitrary $p$ in the limit $N\rightarrow\infty$. Thus, the
generating-function approach combined with saddle-point analysis yields an
exact microcanonical expression for the entropy, from which the remaining
thermodynamic quantities follow directly. In particular, the inverse
temperature is%

\begin{equation}
\beta(u,p)=\frac{1}{\varepsilon k_{B}}\frac{ds(u,p)}{du}=-\frac{1}%
{\varepsilon}\ln x(u).\label{23}%
\end{equation}
This relation also makes transparent the connection with the canonical
treatment, under the formal identification $x=e^{-\beta\varepsilon}$. In this
way, the present microcanonical formulation recovers the corresponding
canonical thermodynamic structure in the thermodynamic limit.

The infinite-temperature point therefore corresponds to $x=1$. Substituting
this value into Eq. (\ref{21}), one obtains%

\begin{equation}
u_{c}=\frac{p-1}{2}.\label{24}%
\end{equation}

For arbitrary finite $p$, this is the energy at which the entropy reaches its
maximum and the inverse temperature vanishes. At the same time, it identifies
the change in the physical saddle-point branch: one has $0<x(u)<1$ for
$u<u_{c}$, $x(u)=1$ at $u=u_{c}$, and $x(u)>1$ for $u>u_{c}$. Accordingly,
$u_{c}$ separates the positive- and negative-temperature sectors of the
microcanonical description. The general solution therefore makes explicit both
the location of the entropy maximum and the way in which the thermodynamic
structure of the model depends on the number of accessible energy levels.

An important consistency test of the formulation is that it reproduces the
known limiting cases in a unified way. For $p=2$, one recovers the standard
two-level expressions,%

\begin{equation}
x(u)=\frac{u}{1-u}\text{, \ \ \ \ \ \ \ }s(u,2)=k_{B}\left[  -u\ln(u)-\left(
1-u\right)  \ln\left(  1-u\right)  \right]  .\label{25}%
\end{equation}
and%
\begin{equation}
\beta(u,2)=\frac{1}{\varepsilon}\ln\left(  \frac{1-u}{u}\right)  .\label{26}%
\end{equation}
In this case, $u_{c}=1/2$, so that the temperature is positive for $0\leq
u<1/2$, infinite at $u=1/2$, and negative for $1/2<u\leq1$. Thus, the general
arbitrary$-p$ formalism correctly reduces to the textbook two-level result.

For $p=3$, the saddle-point equation takes the form%

\begin{equation}
u=\frac{x\left(  1+2x\right)  }{1+x+x^{2}},\label{27}%
\end{equation}
whose physically relevant solution is%
\begin{equation}
x(u)=\frac{u-1+\sqrt{1+6u-3u^{2}}}{2(2-u)}.\label{28}%
\end{equation}
It follows that%

\begin{equation}
s(u,3)=k_{B}\left[  \ln\left(  1+x+x^{2}\right)  -u\ln x\right]  ,\text{
\ \ }x=x(u),\label{29}%
\end{equation}
and%
\begin{equation}
\beta(u,3)=-\frac{1}{\varepsilon}\ln\left[  \frac{u-1+\sqrt{1+6u-3u^{2}}%
}{2(2-u)}\right]  .\label{30}%
\end{equation}
Here $u_{c}=1$, so that the temperature is positive for $0\leq u<1$, infinite
at $u=1$, and negative for $1<u\leq2$. This case is especially relevant
because, among finite nontrivial values of $p$, it is the one for which an
explicit thermodynamic treatment appears to be available in the
literature\cite{liao}. Our result coincides exactly with that case, but it is
obtained here from a different and more general perspective: rather than
treating $p=3$ as an isolated problem, we derive it as a particular
realization of a generating-function formulation that extends directly to
arbitrary $p$. The agreement with the three-level result therefore provides a
nontrivial validation of the present approach and, at the same time,
highlights its broader scope.

The limit $p\rightarrow\infty$ is particularly instructive, because it
connects the present bounded-spectrum problem to the unbounded case. In this
limit, the upper bound of the spectrum is removed, and Eq. (\ref{21}) reduces
to%
\[
u=\frac{x}{1-x},
\]
since $x^{p}\rightarrow0$ for the physically relevant solution $0\leq x<1$.
Hence,%
\begin{equation}
x(u)=\frac{u}{1+u}.\label{31}%
\end{equation}

Substituting Eq. (\ref{31}) into Eq. (\ref{22}), one obtains%
\begin{equation}
s(u,\infty)=k_{B}\left[  \left(  1+u\right)  \ln\left(  1+u\right)  -u\ln
u\right]  ,\label{32}%
\end{equation}
while Eq. (\ref{23}) gives%
\begin{equation}
\beta(u,\infty)=\frac{1}{\varepsilon}\ln\left(  \frac{1+u}{u}\right)
.\label{33}%
\end{equation}
Therefore, in the limit $p\rightarrow\infty$, the inverse temperature remains
positive for all $u\geq0$, and the negative-temperature branch disappears.
This behavior is fully consistent with the loss of the upper bound in the
energy spectrum. More importantly, it shows that the present formulation not
only reproduces isolated special cases, but also captures the full crossover
from finite bounded spectra, where the entropy has a maximum, to the
infinite-level limit, where this maximum is lost. This limiting behavior
reinforces the generality of the solution and clarifies the role played by the
number of accessible levels in the microcanonical thermodynamics of equispaced spectra.%

\begin{figure}[t]
\centering
\includegraphics[width=0.82\textwidth]{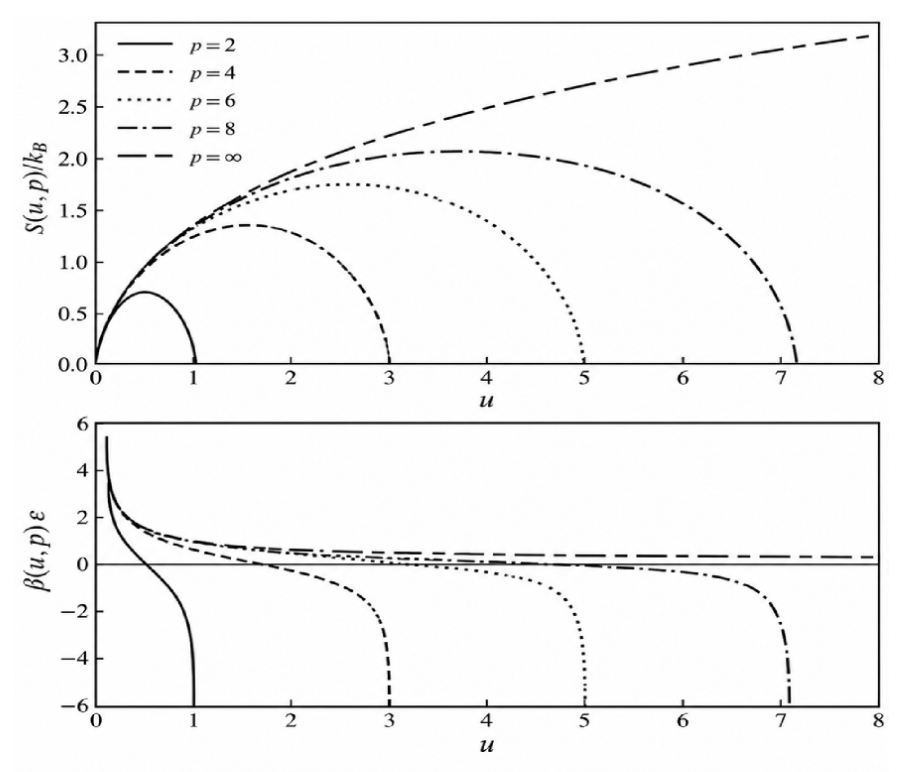}
\caption{Dimensionless entropy per particle, $s(u,p)/k_{B}$ (upper panel), and dimensionless inverse temperature, $\beta(u,p)\varepsilon$ (lower panel), as functions of the reduced energy $u=E/(N\varepsilon)$, for $p=2,4,6,8$, and $\infty$.}
\label{fig:entropy_beta_p}
\end{figure}

To illustrate the thermodynamic content of the general solution, Fig. 1 shows
the dimensionless entropy per particle, $s(u,p)/k_{B}$ (upper panel), and the
dimensionless inverse temperature, $\beta(u,p)\varepsilon$ (lower panel), as
functions of the reduced energy $u=E/(N\varepsilon)$, for $p=2,4,6,8$, and
$\infty$. For every finite $p$, the entropy maximum occurs at $u_{c}=(p-1)/2$,
where the inverse temperature vanishes. As $p$ increases, the entropy maximum
shifts to higher energies and the curve broadens, reflecting the enlargement
of the accessible energy range. The figure thus makes explicit how the
microcanonical thermodynamics evolves with the number of available levels.

In this Letter, we have presented an exact microcanonical formulation, in the
thermodynamic limit, for systems with $p$ equally spaced energy levels. By
combining a generating-function representation of the microcanonical
multiplicity with saddle-point analysis, we obtained closed expressions for
the entropy per particle and the inverse temperature, thereby establishing a
unified description valid for arbitrary $p$. Within this framework, the
entropy maximum is found explicitly at
\[
u_{c}=\frac{p-1}{2},
\]
where the inverse temperature vanishes and changes sign for every finite $p$.

The formulation reproduces in a natural and consistent way the known limiting
cases $p=2$, $p=3$, and $p\to\infty$, while extending them to a single general
solution for arbitrary $p$. It also makes transparent how the thermodynamics
evolves with the number of accessible levels: for finite $p$, the bounded
spectrum leads to an entropy maximum at finite energy and hence to a
negative-temperature branch, whereas in the limit $p\to\infty$ the upper
spectral bound is lost and this structure disappears.

An important aspect of the present result is that the formalism is not
restricted to the spectrum introduced in Eq.~(\ref{1}). It applies immediately
to a spin-$S$ Zeeman multiplet\cite{r1}, since the spectrum $E_{m}%
=-m\varepsilon$, with $m=-S,-S+1,\ldots,S$, differs only by an additive shift
of the energy zero. In this case, the same entropy formula follows after the
replacements $p\rightarrow2S+1$ and $u\rightarrow u+S$, yielding an exact
thermodynamic-limit microcanonical description for arbitrary-spin systems with
bounded spectra.

More broadly, the present methodology provides a useful framework for other
bounded discrete systems whose microcanonical multiplicities admit a
generating-function representation. Possible extensions include isolated
quantum systems with finite local spectra\cite{r5}, as well as discrete
polymer-chain and helix-coil models of statistical mechanic \cite{r7,r8}. The
corresponding finite-$N$ microcanonical thermodynamics, although also of
interest, lies beyond the scope of the present Letter.

\end{document}